\documentclass[letterpaper]{article}
\usepackage{aaai2026}
\usepackage{times}
\usepackage{helvet}
\usepackage{courier}
\usepackage[hyphens]{url}
\usepackage{graphicx}
\usepackage{natbib}
\usepackage{caption}
\usepackage{booktabs}
\usepackage{multirow}
\usepackage{amsmath}
\usepackage{amssymb}
\usepackage{algorithm}
\usepackage{algorithmic}
\usepackage{xcolor}

\title{
Hierarchical Pedagogical Oversight: A Multi-Agent Adversarial Framework for Reliable AI Tutoring 
\thanks{Accepted for presentation at the AAAI 2026 EGSAI Community Activity (AAAI 2026).}
}

\author{
Saisab Sadhu\textsuperscript{\rm 1},
Ashim Dhor\textsuperscript{\rm 1}
}

\affiliations{
\textsuperscript{\rm 1}Indian Institute of Science Education and Research Bhopal\\
Bhopal, Madhya Pradesh, India\\
sadhusaisab@gmail.com, ashimdhor2003@gmail.com
}

\begin{document}

\maketitle

\begin{abstract}
Large Language Models (LLMs) are increasingly deployed as automated tutors to address educator shortages; however, they often fail at pedagogical reasoning, frequently validating incorrect student solutions (sycophancy) or providing overly direct answers that hinder learning. We introduce Hierarchical Pedagogical Oversight (HPO), a framework that adapts structured adversarial synthesis to educational assessment. Unlike cooperative multi-agent systems that often drift toward superficial consensus, HPO enforces a dialectical separation of concerns: specialist agents first distill dialogue context, which then grounds a moderated, five-act debate between opposing pedagogical critics. We evaluate this framework on the MRBench dataset of 1,214 middle-school mathematics dialogues. Our 8B-parameter model achieves a Macro F1 of 0.845, outperforming GPT-4o (0.812) by 3.3\% while using 20× fewer parameters. These results establish adversarial reasoning as a critical mechanism for deploying reliable, low-compute pedagogical oversight in resource-constrained environments.
\end{abstract}

\section{Introduction}
The deployment of Large Language Models (LLMs) as automated tutors offers a scalable solution to the global shortage of qualified educators \cite{kochmar2025bea}. However, recent benchmarks reveal a fundamental reliability gap: LLMs frequently validate incorrect student reasoning to maintain conversational rapport, a phenomenon known as sycophancy \cite{wang2024sycophancy} or fail to identify implicit conceptual errors \cite{demszky2023does}. In educational settings without human-in-the-loop supervision, such ``pedagogical hallucinations" \cite{10.1145/3571730} can actively reinforce misconceptions. These failures often stem from a structural flaw in current agent design: the conflation of \emph{generation} and \emph{evaluation}. Tasking a single model to simultaneously teach and assess its own pedagogical quality creates vulnerability to confirmation bias.

Building on our prior work in financial NLP, where structured adversarial synthesis (SAS) mitigated bias in market analysis \cite{sadhu2025structured}, we adapt this dialectical approach to education with critical architectural modifications. Unlike generative financial synthesis, pedagogical oversight is a constrained classification task requiring strict adherence to instructional taxonomies (Mistake Identification and Guidance Quality).

We propose Hierarchical Pedagogical Oversight (HPO), a framework designed to operationalize adversarial reasoning for education, leveraging principles of AI feedback \cite{bai2022constitutionalaiharmlessnessai} to audit model behavior. HPO decouples the tutoring process from evaluative judgment through a three-phase pipeline: (1) Intelligence Distillation, (2) Adversarial Debate, and (3) Synthesis. We empirically validate HPO on the MRBench dataset \cite{maurya2025mrbench}. Our results show that an 8B-parameter model, when structured via HPO, outperforms GPT-4o by 3.3\% in Macro F1. We demonstrate that this performance is driven by the \textit{adversarial protocol} itself; removing the "Devil's Advocate" moderator results in a significant performance drop.

\section{Methodology: The HPO Framework}

\subsection{Problem Formalization}
We define pedagogical oversight as a classification task. Given a dialogue history $D$, a new student utterance $u_{n+1}$ containing a potential misconception, and a candidate tutor response $R_{\text{cand}}$, the system maps the tuple $(D, u_{n+1}, R_{\text{cand}})$ to a label vector $(y_{\text{MI}}, y_{\text{PG}})$. Here, $y_{\text{MI}} \in \{0,1\}$ denotes Mistake Identification, and $y_{\text{PG}} \in \{0,1,2\}$ denotes Guidance Quality (Direct Solution vs. Partial vs. Effective Scaffolding).

\subsection{Phase 1: Intelligence Distillation}

To ground the downstream debate, three parallel specialist agents (Conceptual Analyst, Behavioral Analyst, Trajectory Analyst) distill the raw dialogue into a "Pedagogical Briefing" ($B$), extracting the mathematical concept, engagement signals, and longitudinal understanding trajectory (see Appendix A). This grounded context prevents hallucination of student intent.

\subsection{Phase 2: Structured Adversarial Debate}

The core of HPO is a deterministic, five-act debate protocol designed to stress-test the candidate response. In \textbf{Act I (Opening)}, a Permissive Critic and a Strict Critic generate opposing theses regarding the response's quality. For instance, given a tutor hint ``Think about what happens when you add fractions,'' the Permissive Critic might argue this provides effective scaffolding, while the Strict Critic contends the hint is too vague to address a specific error (e.g., adding numerators directly). In \textbf{Act II (Cross-Examination)}, a \textbf{Devil's Advocate} agent generates sharp, evidence-based challenges targeting logical gaps in both theses (e.g., ``How do you know the student will interpret this hint correctly given their prior confusion?''). In \textbf{Act III (Rebuttal)}, critics must revise their positions to address these challenges. If defenses remain unconvincing, the Devil's Advocate issues a final ``Press'' in \textbf{Act IV}. Finally, \textbf{Act V} concludes with synthesized summaries. This structure forces deeper reasoning than simple voting.

\subsection{Phase 3: Synthesis and Judgment}
The debate transcript is processed by a final three-agent
pipeline: a Judge adjudicates the winner based on evidence;
a Stress Analyst identifies the remaining vulnerability in the
winning thesis; and a Lead Evaluator synthesizes all inputs
into final classification labels. This hierarchical synthesis prevents the system from overfitting to either critic's initial position.

\subsection{Implementation}
We utilize the AutoGen framework \cite{wu2024autogen} with a Llama-3-8B-Instruct backbone. To adapt the system to the MRBench schema, we apply QLoRA fine-tuning \cite{dettmers2023qlora} specifically to the Lead Evaluator agent (4-bit NF4 quantization, LoRA rank 16, alpha 32).

\section{Experiments}

\subsection{Dataset and Baselines}
We evaluate on MRBench \cite{maurya2025mrbench}, comprising 1,214 test dialogues of middle-school mathematics. We compare HPO against five architectures: S1 (Single-Agent) using Llama-3-70B; S2 (Cooperative) where agents collaborate without debate; S3 (Unstructured Adversarial) involving a simple unmoderated argument; S4 (HPO-Base) using frozen agents; and S5 (HPO-FT) using our full fine-tuned framework. We also benchmark against GPT-4o (Zero-shot) as a gold-standard reference. (See Appendix D for additional ensemble baselines).

\subsection{Main Results}
Table \ref{tab:main} presents the performance on the MRBench test set. Statistical significance was determined via bootstrap resampling ($n=10,000$).

\begin{table}[h]
\centering
\small
\begin{tabular}{lccccc}
\toprule
\multirow{2}{*}{\textbf{System}} & \multicolumn{2}{c}{\textbf{Mistake ID}} & \multicolumn{2}{c}{\textbf{Guidance}} & \textbf{Macro} \\
 & Acc & F1 & Acc & F1 & \textbf{F1} \\
\midrule
GPT-4o & 0.88 & 0.82 & 0.85 & 0.80 & 0.812 \\
Llama-70B & 0.85 & 0.78 & 0.81 & 0.74 & 0.760 \\
\midrule
S1: Single & 0.79 & 0.71 & 0.76 & 0.68 & 0.695 \\
S2: Cooperative & 0.86 & 0.80 & 0.83 & 0.77 & 0.785 \\
S3: Unstructured & 0.88 & 0.82 & 0.85 & 0.78 & 0.800 \\
S4: HPO-Base & 0.90 & 0.84 & 0.87 & 0.81 & 0.825 \\
\textbf{S5: HPO-FT} & \textbf{0.91} & \textbf{0.86}$^*$ & \textbf{0.89} & \textbf{0.83}$^*$ & \textbf{0.845}$^*$ \\
\bottomrule
\end{tabular}
\caption{Performance comparison on MRBench (Macro F1). S5 (HPO-FT) outperforms both the cooperative baseline and the closed-source GPT-4o model. $^*$ indicates statistical significance ($p < 0.01$) vs. GPT-4o via bootstrap resampling ($n=10{,}000$, 95\% CI $\pm 0.015$).}
\label{tab:main}
\end{table}

\textbf{Impact of Adversarial Structure:} S4 (HPO-Base) outperforms S2 (Cooperative) by +4.0\% F1. This confirms that the adversarial process generates higher-fidelity signal than mere cooperation. Furthermore, S4 outperforms S3 (Unstructured) by +2.5\%, validating the necessity of the "Devil's Advocate" moderator to prevent superficial consensus. Notably, our S5 system (8B parameters) outperforms GPT-4o (175B+) by +3.3\%, suggesting that for specialized evaluative tasks, a structured agentic workflow is superior to raw model scale.

\subsection{Ablation Study}
To isolate the source of gains, we ablated components of S5 (Table \ref{tab:ablation}). The removal of Phase 1 Distillation causes the largest drop (-8.3\%), highlighting the importance of grounding the debate. Notably, removing the Devil's Advocate (-4.2\%) hurts performance more than removing fine-tuning (-2.0\%), reinforcing our hypothesis that the \textit{interaction structure} is more critical than the model weights.

\begin{table}[h]
\centering
\small
\begin{tabular}{lcc}
\toprule
\textbf{Configuration} & \textbf{Macro F1} & \textbf{$\Delta$} \\
\midrule
\textbf{Full HPO-FT} & \textbf{0.845} & - \\
(-) Remove Phase 1 Distillation & 0.762 & -0.083 \\
(-) Remove Devil's Advocate & 0.803 & -0.042 \\
(-) Remove Multi-Turn Protocol & 0.815 & -0.030 \\
(-) Remove QLoRA Fine-Tuning & 0.825 & -0.020 \\
\bottomrule
\end{tabular}
\caption{Ablation study on the HPO-FT architecture.}
\label{tab:ablation}
\end{table}

\section{Discussion and Conclusion}

We introduced Hierarchical Pedagogical Oversight (HPO), operationalizing structured adversarial synthesis for automated tutoring reliability. HPO mitigates "mode collapse" where multi-agent systems converge on sycophantic consensus, instead exploring probability distribution tails to interrogate whether polite responses conceal pedagogical harm.

Adapting principles from our financial analysis work, HPO implements a hybrid architecture: generalist agents for dialectical exploration, fine-tuned specialists for schema-constrained output. Demonstrating that 8B-parameter models outperform proprietary systems establishes a pathway for "Pedagogical Safety Layers" in resource-constrained environments (rural schools, low-bandwidth settings). This confirms interaction structure, not model scale, is the bottleneck for reliable AI tutoring.

\textbf{Limitations:} Our evaluation focuses on middle-school mathematics; cross-domain validation is needed. Multi-turn debate incurs $\sim$4.2s latency, suitable for asynchronous grading rather than real-time intervention. Future work includes deployment validation and adaptive debate mechanisms.

\bibliography{aaai2026}
%

\appendix

\section{Appendix A: Specialist Agent Definitions}
Phase 1 employs three specialist agents to distill the dialogue context. These agents are prompted with specific personas to ensure diverse information extraction:
\begin{itemize}
    \item \textbf{Conceptual Analyst:} ``You are an expert Mathematics Curriculum Designer. Your task is to identify the specific mathematical concept in the dialogue and pinpoint the precise nature of the student's error (e.g., calculation error vs. conceptual misunderstanding). Ignore tone; focus only on the logic.''
    \item \textbf{Behavioral Analyst:} ``You are an Educational Psychologist. Analyze the student's engagement signals. Are they frustrated, over-confident, or guessing? Assess the Tutor's tone: is it encouraging, dismissive, or robotic?''
    \item \textbf{Trajectory Analyst:} ``You are a Learning Trajectory Specialist. Trace the trajectory of understanding across the previous $k=5$ turns. Is the student repeating a previously corrected error? Has the tutor's previous scaffolding failed? Determine if the student is progressing toward the solution or regressing into confusion.''
\end{itemize}

\vspace{0.5em}
\noindent \textbf{Example Distillation Output:}

Given a dialogue where a student incorrectly adds fractions as $\frac{1}{2} + \frac{1}{3} = \frac{2}{5}$, the agents produce:

\begin{itemize}
    \item \textbf{Conceptual Analyst:} ``Mathematical Concept: Fraction addition with unlike denominators. Error Type: Conceptual misunderstanding---student is adding numerators and denominators directly, violating the need for common denominators.''
    \item \textbf{Behavioral Analyst:} ``Engagement: Student appears confident (uses definitive language: `I got 2/5'). Tutor tone: Neutral, asks `Are you sure?' without elaboration.''
    \item \textbf{Trajectory Analyst:} ``Over past 5 turns, student has successfully solved fraction addition with like denominators (2/5 + 1/5). This suggests procedural knowledge exists but hasn't generalized to unlike denominators. No prior correction of this specific error.''
\end{itemize}

\section{Appendix B: Adversarial Prompt Engineering}
\textbf{Devil's Advocate (Exact System Prompt):}
\begin{quote}
\small
\texttt{ROLE: You are the Devil's Advocate in a pedagogical debate.}

\texttt{TASK: Analyze arguments from both the Permissive and Strict critics. Identify 1-3 logical gaps or unsupported assumptions in EACH argument.}

\texttt{OUTPUT FORMAT:}\\
\texttt{- Challenge to Permissive Critic: [specific logical gap]}\\
\texttt{- Challenge to Strict Critic: [specific logical gap]}

\texttt{RULES:}\\
\texttt{1. Be precise. Avoid generic critiques.}\\
\texttt{2. Force analysts to defend reasoning with evidence from dialogue.}\\
\texttt{3. If an argument assumes student mental state, ask: what supports this?}
\end{quote}

\vspace{0.5em}
\noindent \textbf{Lead Evaluator (Exact System Prompt):}
\begin{quote}
\small
\texttt{ROLE: You are the Lead Evaluator synthesizing a pedagogical debate.}

\texttt{TASK: Assign final labels based on debate winner and stress analysis.}

\texttt{OUTPUT (JSON):}\\
\texttt{\{"mistake\_identified": true/false,}\\
\texttt{ "guidance\_quality": 0/1/2,}\\
\texttt{ "confidence": 0.0-1.0,}\\
\texttt{ "reasoning": "brief justification"\}}

\texttt{LABELS:}\\
\texttt{- mistake\_identified: Did tutor correctly identify student error?}\\
\texttt{- guidance\_quality: 0=Direct answer, 1=Partial hint, 2=Effective scaffolding}
\end{quote}

\section{Appendix C: Detailed Error Analysis}
To understand the limitations of HPO, we analyzed the 15.5\% of cases where the system failed. Table \ref{tab:confusion} shows the confusion matrix for the ``Guidance Quality'' task.

\begin{table}[h]
\centering
\small
\begin{tabular}{c|ccc}
\toprule
\textbf{Actual \textbackslash Pred} & \textbf{0} & \textbf{1} & \textbf{2} \\
\midrule
\textbf{0} & \textbf{382} & 41 & 8 \\
\textbf{1} & 56 & \textbf{298} & 32 \\
\textbf{2} & 12 & 39 & \textbf{346} \\
\bottomrule
\end{tabular}
\caption{Confusion Matrix for Guidance Quality (Predicted vs. Actual). Classes: 0 (No Guidance), 1 (Partial), 2 (Effective).}
\label{tab:confusion}
\end{table}

\noindent \textbf{Failure Modes:}
\begin{enumerate}
    \item \textbf{Ambiguity (Class 1 vs 0):} The most common error (56 cases) is misclassifying ``Partial Guidance'' as ``No Guidance.'' Qualitative review suggests this occurs when the tutor provides a very subtle hint that the Strict Critic dismisses as ``too vague,'' swaying the Judge. \textit{Example:} Tutor says ``Think about the denominator.'' Strict Critic argues this is too generic; Permissive counters it directs attention. Judge sided with Strict, but ground truth labeled it as Partial.
    \item \textbf{Over-Correction:} In 41 cases, the model flagged ``No Guidance'' as ``Partial.'' This typically happens when the tutor is polite but factually unhelpful, and the Permissive Critic successfully argues that ``encouragement is a form of guidance.'' \textit{Example:} Tutor says ``You're doing great! Keep trying.'' Permissive argues emotional support aids learning; Strict argues no content guidance provided. Judge sided with Permissive incorrectly.
    \item \textbf{Effective vs Partial (Class 2 vs 1):} In 39 cases, truly effective scaffolding was underrated as partial. This occurs when the Devil's Advocate challenges whether a hint is ``actionable enough,'' causing the Judge to become overly conservative. Future work should refine the Devil's Advocate's prompting to distinguish between ``vague'' and ``appropriately open-ended'' scaffolding.
\end{enumerate}

\section{Appendix D: Additional Baselines}
To verify that the performance gains stem from the \textit{adversarial structure} and not just the use of multiple agents or sampling, we compared HPO against two strong ensemble baselines using the Llama-3-8B backbone.

\begin{table}[h]
\centering
\small
\begin{tabular}{lc}
\toprule
\textbf{Method} & \textbf{Macro F1} \\
\midrule
Single Agent (Llama-3-8B) & 0.695 \\
Self-Consistency (k=5, Majority Vote) & 0.742 \\
Ensemble (3 Independent Agents) & 0.768 \\
\textbf{HPO-FT (Ours)} & \textbf{0.845} \\
\bottomrule
\end{tabular}
\caption{Comparison against Ensemble Baselines.}
\label{tab:baselines}
\end{table}

\noindent \textbf{Insight:} While Self-Consistency (+4.7\%) and Ensembling (+7.3\%) improve over the single agent, they fail to reach the performance of HPO. This confirms that the \textit{dialectical process} of debate uncovers insights that simple voting or sampling cannot.

\section{Appendix E: Training Details}
The Lead Evaluator was fine-tuned using the \texttt{trl} library on a single NVIDIA A100 (40GB).
\begin{itemize}
    \item \textbf{Base Model:} \texttt{meta-llama/Meta-Llama-3-8B-Instruct}
    \item \textbf{Quantization:} 4-bit NF4 (NormalFloat4)
    \item \textbf{LoRA Config:} Rank $r=16$, Alpha $\alpha=32$, Dropout=0.05
    \item \textbf{Learning Rate:} 2e-4 (Cosine Schedule)
    \item \textbf{Batch Size:} 16 (Gradient Accumulation Steps=2)
    \item \textbf{Epochs:} 3
\end{itemize}

\section{Appendix F: Computational Cost Analysis}
To address deployment feasibility, we provide a breakdown of inference time per evaluation on NVIDIA A100 (40GB):

\begin{table}[h]
\centering
\small
\begin{tabular}{lcc}
\toprule
\textbf{Component} & \textbf{Time (s)} & \textbf{\% Total} \\
\midrule
Phase 1: Distillation (3 parallel agents) & 0.8 & 19\% \\
Phase 2: Debate (5 turns, sequential) & 2.6 & 62\% \\
Phase 3: Synthesis (3 sequential agents) & 0.8 & 19\% \\
\midrule
\textbf{Total HPO-FT} & \textbf{4.2} & \textbf{100\%} \\
\midrule
Single Llama-3-8B call & 0.4 & -- \\
GPT-4o API call (avg) & 1.2 & -- \\
\bottomrule
\end{tabular}
\caption{Inference latency breakdown. HPO is 10$\times$ slower than single-agent but achieves 15\% higher F1, acceptable for batch evaluation (e.g., grading 1000 responses overnight takes $\sim$70 minutes).}
\label{tab:cost}
\end{table}

\noindent \textbf{Cost-Performance Tradeoff:} For real-time tutoring, adaptive mechanisms could reduce latency (e.g., skip Phase 2 if Phase 1 confidence $>$ 0.95). For batch quality assurance, the 4.2s overhead is negligible compared to human grading ($\sim$2-5 minutes per response).

\end{document}